# Physical understanding of the static magnetic field's synergistic enhancement of cold atmospheric pressure plasma treatment

Ramin Mehrabifard[1], Zeinab Kabarkouhi[2], Fatemeh Rezaei[3], Kamal Hajisharifi[1], Hassan Mehdian[1], Eric Robert[4]

[1] Department of Physics and Institute for Plasma Research, Kharazmi University, 49 Dr. Mofatteh Avenue, Tehran, Iran

[2] Laser and Plasma Research Institute, Shahid Beheshti University, Evin, Tehran, Iran

[3] Department of Physics, K. N. Toosi University of Technology, Shariati, Tehran, Iran

[4] GREMI, UMR 7344, CNRS/Université d'Orléans, 45067 Orléans, France

Corresponding author: k.hajisharifi@gmail.com, hajisharifi@khu.ac.ir , fatemehrezaei@kntu.ac.ir

**Abstract**

In the last decades, to improve the CAP treatment efficiency, its biological effects in combination with other physical modalities have widely investigated. However, the physical insight into most of supposed synergistic effects remained elusive. In this regard, the synergetic effect of cold plasma and magnetic field has been used for different applications, especially due to considerable synergistic in biological media reactivity. In the present paper, using a 420 mT N42 magnet, the effect of the perpendicular external static magnetic field (SMF) on the cold atmospheric pressure plasma (CAP) characteristics, such as electron temperature and density, are investigated based on the optical emission spectroscopy, utilizing the Boltzmann plot method, Saha-Boltzmann equation and Specair software simulation. Results showed that the rotational and electronic excitational temperature experienced 100 K and 550 K increases in the presence of SMF, respectively. While the vibrational and translational temperatures remained constant. Moreover, electron temperature estimated as 1.04 eV in the absence of SMF and increased up to 1.24 eV in the presence of SMF. In addition, the Saha-

Boltzmann equation illustrated that the electron density increased in presence of the additional SMF. The results of the present study indicated that the magnetic field could be an assistant to the cold plasma effect, beneficial in medical applications due to modifications in plasma temperature and electron density.

## 1. Introduction

Cold atmospheric pressure plasma (CAP) has gained attractive interest in recent years due to a vast variety of applications such as surface modification [1], medical science [2,3], and material processing [4]. A plasma jet reactor or a typical CAP produces a large number of active species such as electrons, ions, and radicals in a long effluent that extends into the atmosphere. Meanwhile, the gas temperature is somewhat low, equal to the room temperature [5].

Nowadays, much attention has been focused on the improvement of the cold plasma effect, especially for medical applications [6]. One of these methods is an assist by an external static magnetic field (SMF). Generally, SMFs have unique properties such that they are difficult to shield and quickly infiltrate any system. Curative properties of the magnetic field have been seen in various medical applications, including wound healing [7] and bacterial deactivation [8]. For instance, according to Sadri et al. [9], mesenchymal stem cell alignment and proliferation rate are affected by exposure to SMF up to 24 *mT*. Furthermore, the therapeutic effect of the magnetic field on different cancer cells has also been studied in the last decade [10–12].

The synergetic effect of CAP and SMF has been used for different fields of plasma applications, such as antibacterial treatment [13,14]. For instance, Mackinder et al. [15] investigated cold plasma sterilization in the presence of an external magnetic field. They showed that the most efficient sterilization could be reached at an optimal pressure with an external magnetic field. Han et al. [16] improved the plasma effect on water treatment with an external SMF. They concluded that ROS increased under the action of a magnetic field concentration, and melanoma cells' vitality dropped significantly. Recently, Cheng et al. [6] introduced a new effective method by simultaneous usage of a magnet and plasma jet for cancer cells treatment. The combination effects of the magnetic field and cold plasma on water sample in the biomedical applications were studied by Xu et al. in Ref [17]. They discovered that this method increased the ionization collisions and plasma jet discharge intensity. They also showed that, enhancing the transverse magnetic field, the concentration of the aqueous reactive species such as $O_2$, OH, and $H_2O_2$ and the efficiency of inactivating living tissues like cancerous cells and Escherichia coli also increased. In this our work, we showed that the synergetic effect of cold plasma and SMF will improve the plasma treatment effect on breast cancer cells [18]. The results confirmed that the cell viability and migration rate dramatically decreased in the presence of SMF, are in excellent agreement with other literature findings [6,19].

In all previous researches, the external magnetic field effect on CAP treatment efficiency has just been reported biologically and the physical insight into the occurring mechanisms has been missed. In the present study, we will try to fill this gap and show physically why cold plasma is more effective in an external magnetic field. For this purpose, the effect of the external magnetic field on the plasma parameters having essential roles in biological responses of CAP treatment, including plasma temperature, electron density, and electron temperature will be investigated. It should be mentioned that the gas temperature and the concentration of electrons are vital parameters in the treatment process by plasma jet.

## 2. Experimental set-up

The experimental set-up mainly consists of a plasma jet, magnet, calibrated gas dosing valve, Radio Frequency (RF) generator, and the diagnostic system shown in figure 1. The powered electrode is a 1 mm steel wire connected to an RF source (13.56 *MHz*) through a matching impedance network, and the grounded electrode is wrapped around a 5 *mm* glass tube (the glass covered with a Teflon strip). A steady plasma is produced by an atmospheric pressure argon flow with a rate of 650 *SCCM* and 20*W* input power. The applied permanent magnet (IRmagnet, IR) is manufactured from N42 alloy, and its magnitude is measured using a Gauss meter (HT201 Portable Digital Gauss Meter).

Optical emission spectroscopy (OES) is a commonly used method for studying atmospheric pressure discharge and its composition [20]. Moreover, spectroscopy can evaluate the influence of the magnetic field on the mixing ratio of the reactive species in the cold plasma. In this experiment, an Avantes multi-channel fiber-optic spectrometer with a wavelength range of 200–1200 *nm*, and a resolution of

0.19 *nm* is employed to extract plasma parameters. It's worth mentioning that all the spectra are the recorded spectra subtracting background emissions shown in figure 2. Furthermore, the spectrometer is calibrated for absolute irradiance regularly by a Halogen Deuterium light source. Several studies have calculated excitation temperature using the ratio of relative intensities of the atomic lines, Assuming that this value is comparable to the temperature of electrons in the discharge [21]. Another emission spectroscopy application is to estimate the gas temperature by fitting the rotational/vibrational spectra of molecular species like hydroxyl radicals or nitrogen molecules with a simulated spectrum during variation of the assumed system's temperatures [22]. For instance, Bruggeman et al. [23] presented a critical assessment of this approach, and confirmed the precautions for interpreting the calculated temperatures.

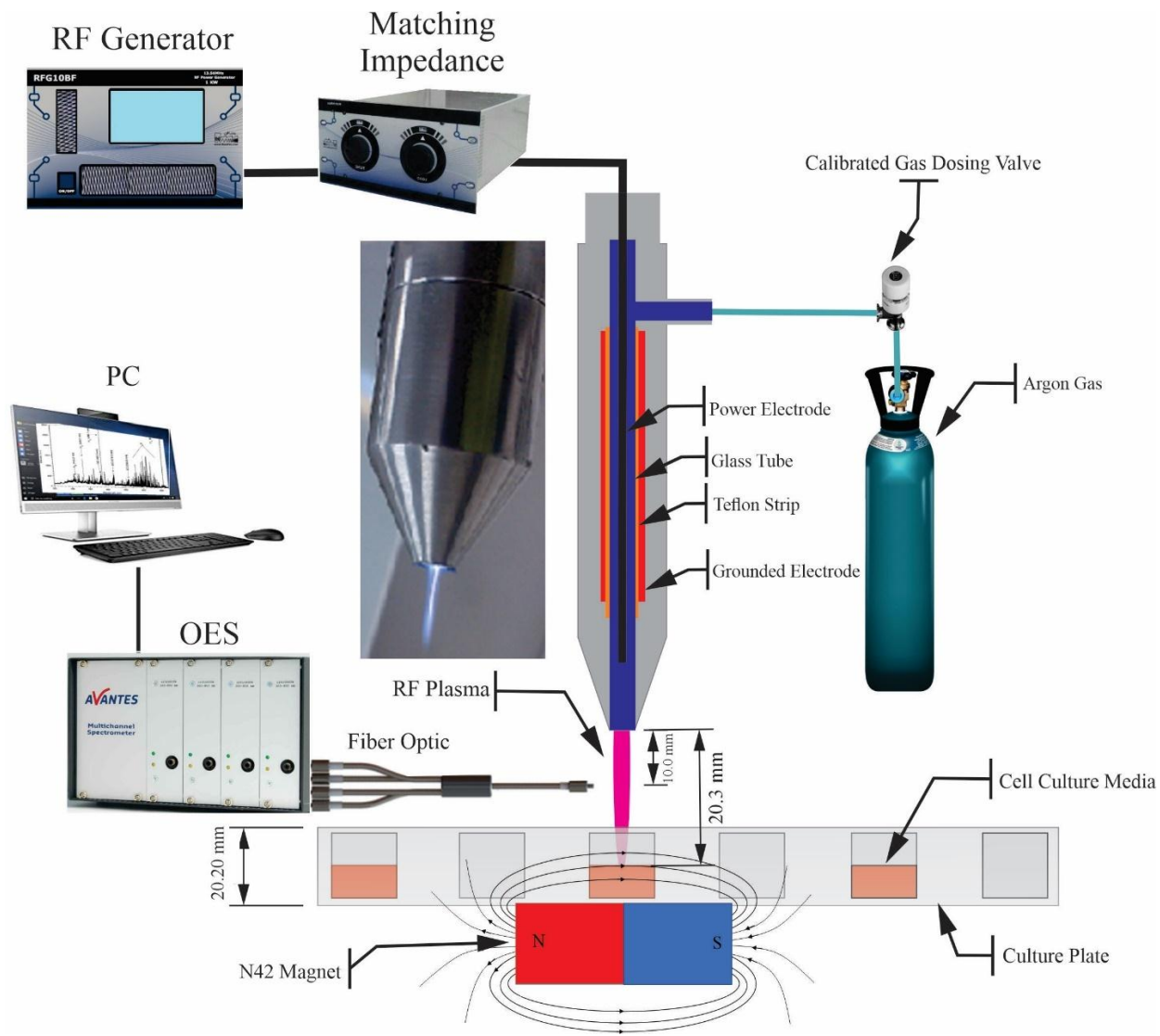

Figure 1. Schematic diagram of RF plasma setup including real nozzle, OES, and a magnet.

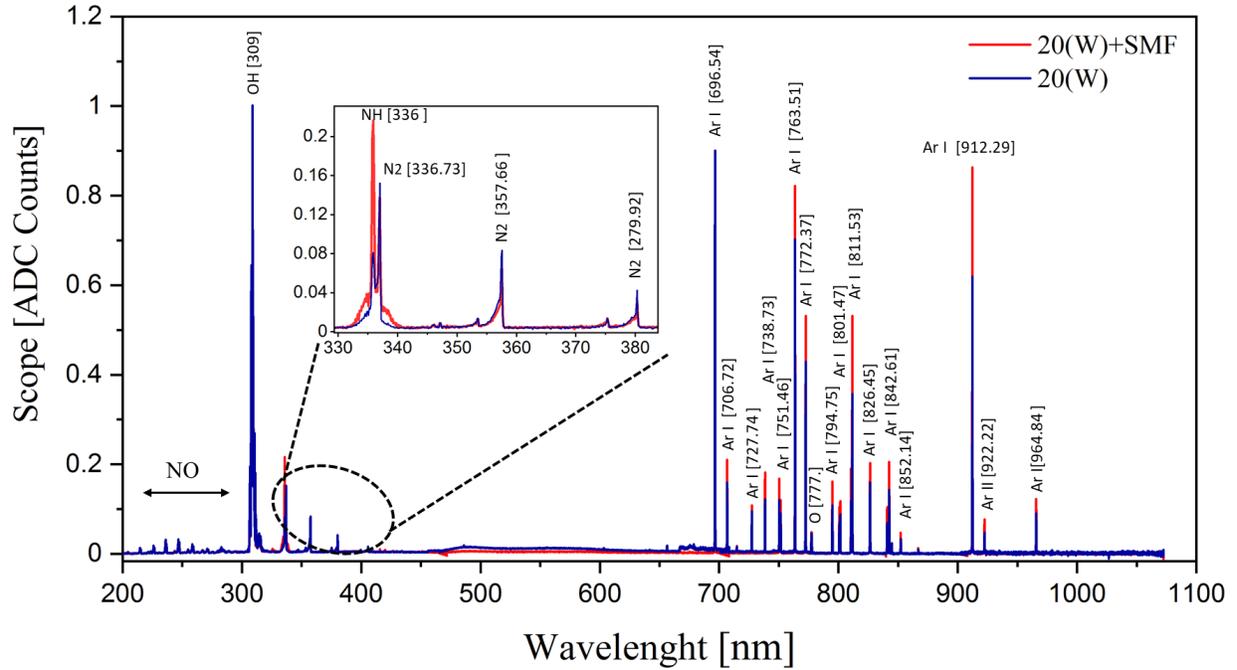

Figure 2. Optical emission measurement in the presence and absence of SMF.

## 3. Theoretical method
### 3.1. Boltzmann method for temperature measurement

In physical study of the CAP, the temperature of electrons is a crucial metric to consider. It should be noted that the principle of local thermodynamic equilibrium (LTE) is an essential criterion in calculating the plasma electron temperature. Generally, the thermodynamic condition for complete thermodynamic equilibrium (CTE) holds for LTE in plasma, but in the absence of CTE, the partial local thermodynamic equilibrium (PLTE) can still be attained. This situation enables one to use the generic thermodynamic equation in constrained conditions. This study investigates the validity requirements for PLTE and LTE conditions in cold plasma emission (RF jet), created to treat cancer cells. It should be mentioned that minimal densities of free electrons are required to enable PLTE and LTE conditions, according to Griem criteria [24]. The following equation represents the electron density requirements for the PLTE and LTE, respectively:

$$N_e \geq 7.4 \times 10^{18} \frac{z^7}{n^{17/2}} \left(\frac{K_B T}{E_x}\right)^{\frac{1}{2}} \ [cm^{-3}] \tag{1}$$

and

$$N_e \geq 9.2 \times 10^{17} z^7 \left(\frac{K_B T}{E_x}\right)^{\frac{1}{2}} \left(\frac{E_2 - E_1}{E_x}\right)^3 \ [cm^{-3}], \tag{2}$$

where z is the effective charge by bound electron, n is the principal quantum number of the lowest level in the partial LTE condition, $K_B$ is the Boltzmann's constant, $T$ is the electron temperature, $E_1$ is lower level, $E_2$ is the upper-level energy, and $E_x$ is the ionization energy. For such plasma systems in LTE, the Boltzmann plot is used to calculate the electron temperature $T_e$ as [25,26]:

$$Ln\left(\frac{I_{ij}\lambda_{ij}}{A_{ij}g_j}\right) = \frac{-E_i}{K_B T_e}, \tag{3}$$

in which $I_{ij}$ is the intensity of the emitting light, $\lambda_{ij}$ is the wavelength, $g_j$ is the statistical weight of the upper level, $A_{ij}$ is the transition probability from level $i$ to level $j$, and $E_i$ is the energy of the upper level and $K_B$ is the Boltzmann's constant.

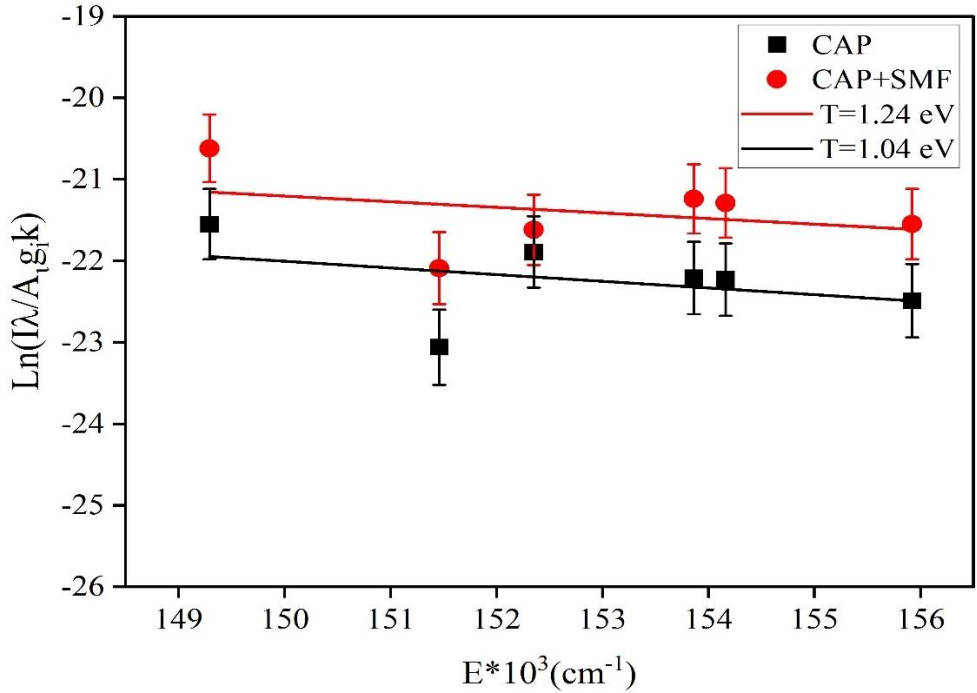

Figure 3 shows the plot of $Ln(I_{ij}\lambda_{ij}/A_{ij}g_j)$ versus $E_k$ for extraction of $T_e$. The excitation temperature $T_{exc}$ is equal to the temperature of the electron $T_e$, in the condition of Local Thermodynamic Equilibrium (LTE).

### 3.2. Saha-Boltzmann equation for density measurement

The Saha-Boltzmann equation is used to calculate the electron density using atomic and ionic spectral lines emitted from the plasma radiation as [27]:

$$n_e = \frac{I_Z^*}{I_{Z+1}^*} 6.04 \times 10^{21} (T)^{3/2} \times \exp[(-E_{k,z+1} + E_{k,z} - \chi_z)/K_B T] \; cm^{-3}, \qquad (4)$$

where $I_Z^* = I_Z \lambda_{ki,z} / g_{k,Z} A_{ki,Z}$ and $\chi_z$ is the ionization energy of the species in the ionization stage. It should be mentioned that the reduction of the ionization energy due to the plasma interactions is negligible. Hence, it can be ignored from equation (4). To obtain the value of $n_e$, it is better to average the electron density extracted from the intensity ratio of Ar I lines to Ar II lines (912 *nm* Ar I, 922 *nm* Ar II) (738 *nm* Ar I, 922 *nm* Ar II).

### 3.3. Vibrational, translational, and rotational temperatures

Molecules are abundant in atmospheric pressure plasmas and significantly contribute to the emission spectrum. The molecules' rotational, vibrational, and electronic excitational temperatures can be calculated from emission spectra providing significant information about the intermolecular interaction, i.e., plasma conditions [28]. These molecular temperatures are related to the equilibrium assumption between the corresponding levels. For instance, the rotational temperature can be obtained from the Boltzmann distribution of the rotational levels. However, when the atmospheric pressure plasma is out of equilibrium, different species will have different temperatures. Generally, light electrons acquire their translational energy directly from the electric field, so the electron temperature is significantly higher than heavy particles temperature known as the gas temperature. More complex interactions exist between different species and also there are various degrees of freedom in molecules. The rotational temperature of the diatomic molecules in the plasma has been supposed to be equal to gas translational temperature due to the fast rotational-translational equilibrium in atmospheric pressure plasmas [29–31]. But the assumption should be considered with special care because the collisions might not be enough to equilibrate the rotational distribution of levels during gas translational motion [32,33].

In this research, the Second Positive System (SPS) spectrum of the nitrogen molecule is used for temperature measurements. Positive here means the neutral molecule [34] and the second term represents their occurrence condition [35]. The SPS peaks group originates from the transitions between different vibrational levels of two excited electronic levels (B, C) [36]. To be more precise, the *v* quantum number indicates the molecule's vibrational motion as a harmonic oscillator. A classification of transitions between electronic-vibrational-rotational levels is performed by the vibrational number change $\Delta v$, which is based on neighboring location of similar $\Delta v$ transitions on the spectrum. In this paper, the rotational and vibrational temperatures of nitrogen molecules estimated by Specair software. Specair by Laux C.O. is a well-known FORTRAN program which depict the air spectra [29,37,38]. This program was developed from the NEQUIR code by NASA [39]. Here, the convenient $\Delta v = -1$ range of the vibrational transition of $N_2(C^3\Pi_u - B^3\Pi_g)$ spectrum is simulated by Specair to get an estimation of the rotational temperature and possibly gas temperature. Generally, a normalization of the Specair and recorded spectrum to a maximum value of 1 is necessary for simulations.

## 4. Result

In this research, OES is used to analyze cold plasma characteristics. Figure 2 shows the spectral emission of the argon plasma in the presence and absence of SMF between the wavelength ranges of 100 to 1100 *nm*. As clearly seen in this figure, the emission intensity increases at most wavelengths in the presence of the magnetic field.

The calculation indicates that the electron temperature $T_e$ and electron density $n_e$ is estimated as 1.04 *eV* and $4.65 \times 10^{15}\, cm^{-3}$ for the absence of SMF, and 1.24 *eV* and $4.095 \times 10^{16} cm^{-3}$ for the presence of SMF, respectively. On the other hand, the PLTE and LTE electron density criteria are evaluated and their magnitudes are estimated as PLTE=$1.35 \times 10^{11} cm^{-3}$ and LTE=$2.67 \times 10^{12} cm^{-3}$, respectively. Therefore, these results illustrate that the RF plasma meets both of PLTE and LTE criteria.

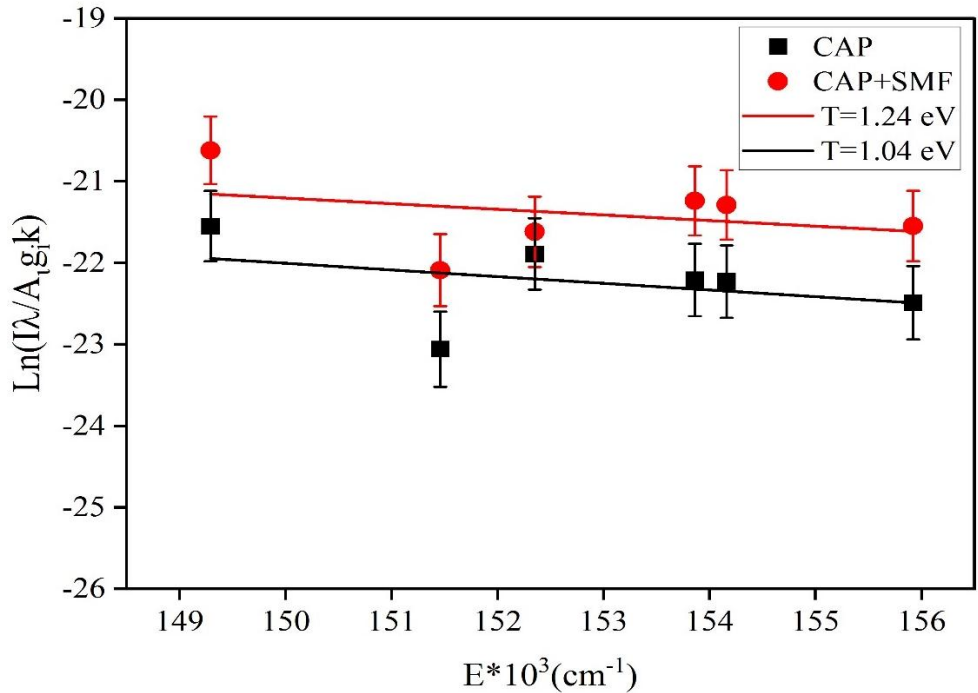

Figure 3. Boltzmann plot for calculation of the electron temperature.

The recorded spectrum in the wavelength interval of 305-385 *nm* is shown in Figure 4 which exhibits reactive oxygen and nitrogen species, including $N_2$, OH, NO, and nitrogen SPS. The nitrogen originating from the ambient air constitutes a considerable percentage of the atmospheric pressure plasma's emission. The bands of SPS are designated in the plot which manifests their importance. Clearly, the discharge cannot ionize the $N_2$ molecules into the nitrogen ion molecule $N_2^+$.

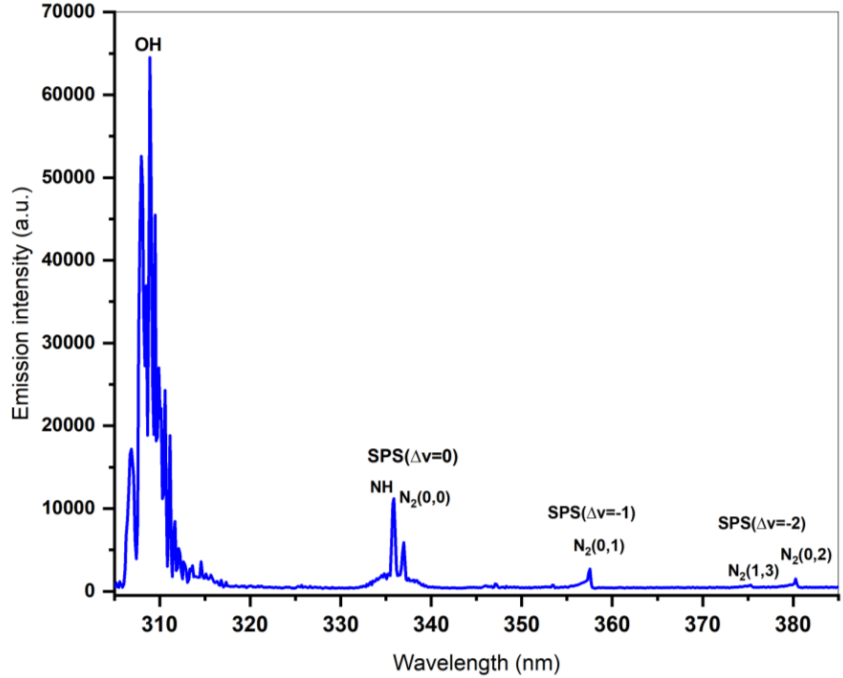

Figure 4. The spectrum of CAP in spectral ranges of 305-385 nm.

A comparison of the spectrum concerning the magnetic field presence is depicted in Figure 5. As can be seen in this figure, the magnetic field only affects the OH transitions through augmented intensities. The selective impact on the limited range can be attributed to the higher intensities of these transitions which pronouncedly exhibit the field effects.

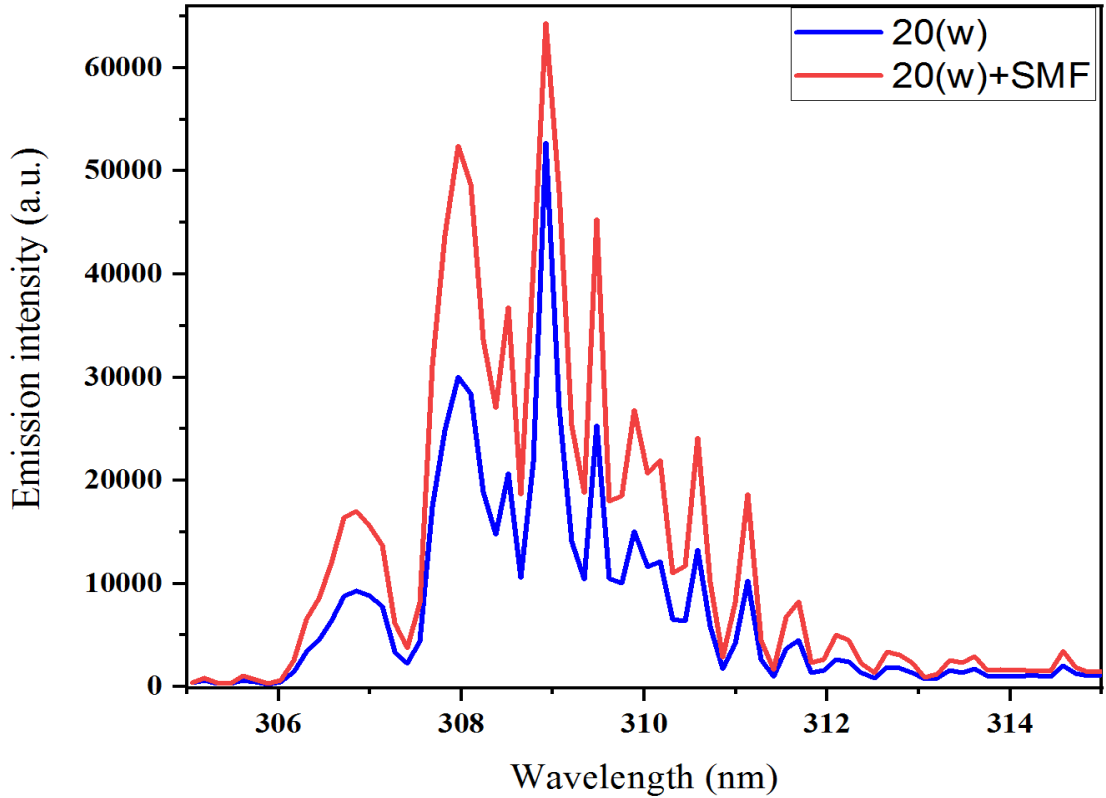

Figure 5. Comparison of the spectra for two cases of with and without presence of magnetic field, by enlarging the affected range.

In the next step, Specair simulations have been employed to determine the temperatures in the cases of with and without the static magnetic field. Figure 6 indicates that translational and vibrational temperatures were equal and about 1800 K regardless of magnetic field existence, suggesting holding the vibrational-translational equilibrium condition. This is not an expected situation for atmospheric pressure plasma where the rotational-translational equilibrium is much more frequent.

As it is clearly seen in figure 6, the magnetic field has increased the rotational temperature slightly from 500 K to 600 K. The obtained temperatures are far from a cold treatment condition required for plasmas applying to cells. Moreover, the result indicates that the estimated temperatures do not represent the gas temperature. It's been suggested that large energy transfer from electrons to the rotational populations in argon plasmas will provide a high $T_{rot}$ [40]. Thus, the equality is not an explicit assumption.

Overall, only rotational and electronic excitational temperatures were affected by the magnetic field application.

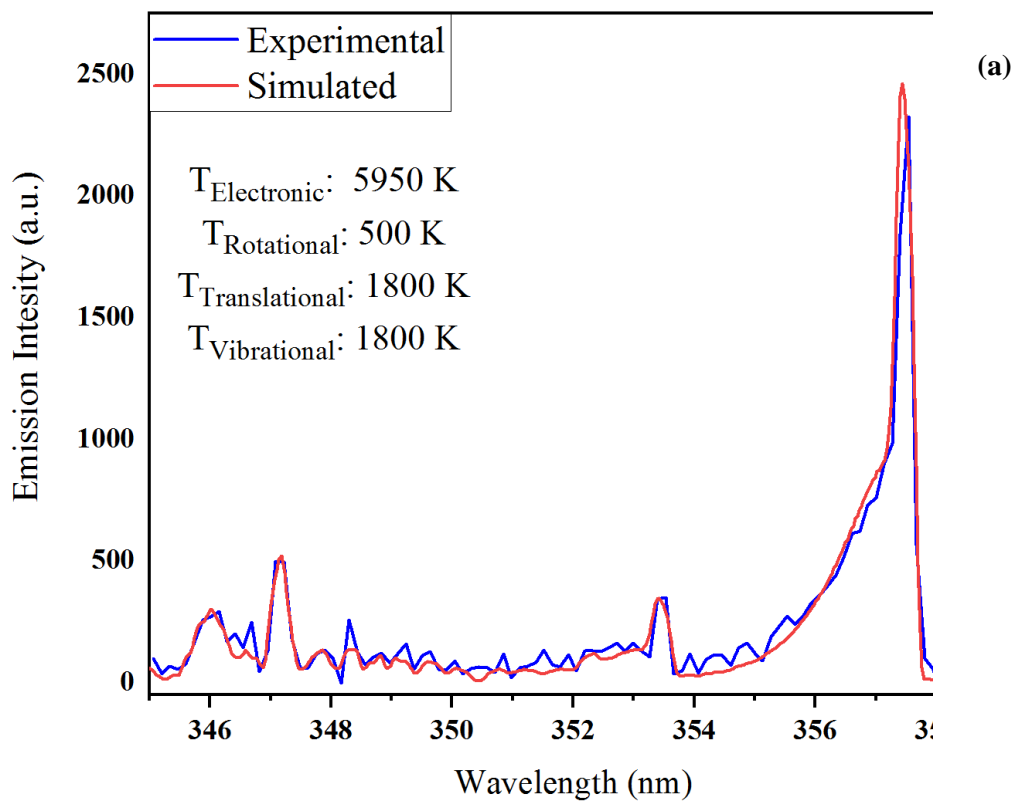

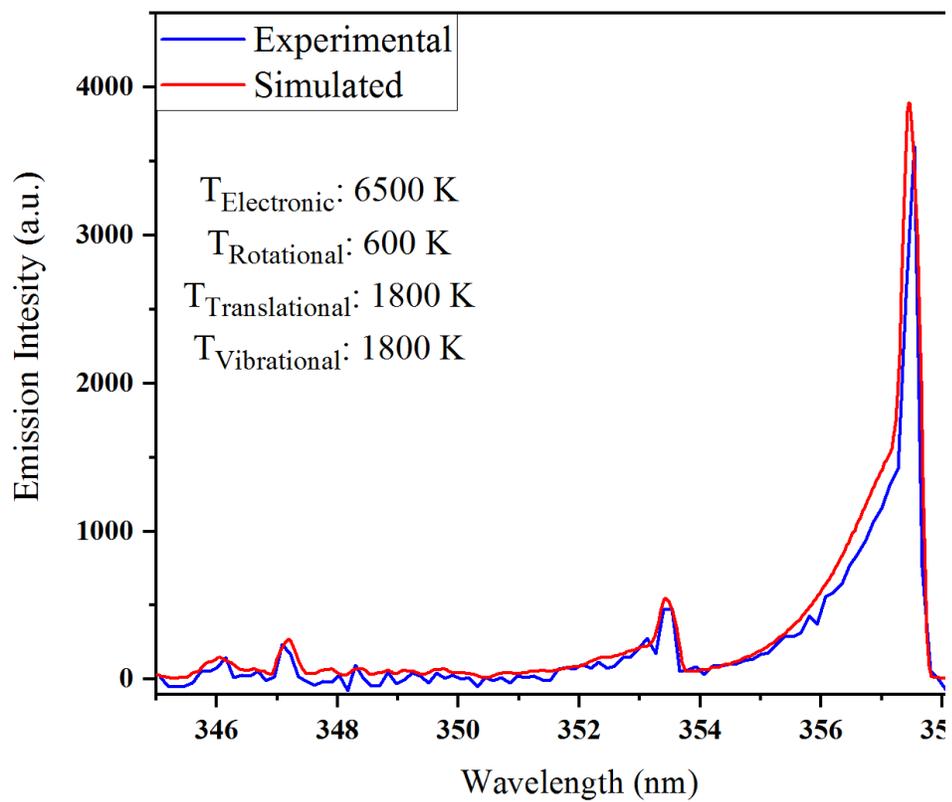

**(b)**

Figure 6. The simulated versus experimental spectrum; (a). In the absence of the magnetic field, (b), and in the presence of the magnetic field.

## 5. Discussion

In the last years, to improve CAP treatment efficiency, its biological effects in combination with other physical modalities have been widely investigated. However, the physical insight into the most synergic effects remained elusive. In this regard, the synergic effect of cold atmospheric plasma and the external magnetic field has attracted lots of attention, due to the considerable synergistic enhancement in biological mediums. Generally, this efficiency improvement may arise from the effect of the magnetic field on the generated plasma, on the target or cell, or the simultaneous effect of the magnetic field and plasma on the target [6,18,19]. Regarding the idea that ROS and RNS may play a critical role in plasma interactions with tissue or cells, the synergic effects may be accompanied by increasing reactive species production in the system. Besides the direct roles of the reactive species, recent researches have proposed electrons as important short-term species with a great impact on plasma characteristics. The electron is considered one of the key plasma components, and any change in its temperature and density influences many parameters and behavior of the system. Generally, increasing the number density of electrons in the plasma can significantly affect the biological target which can make the treatment process faster. Furthermore, with the increase of the electron density, active species will show a good rising [41]. In many cancer cell treatments, including ovaries, brain tumors, and skin cancers with plasmas comprising higher electron densities, the cold plasma performed much better than conventional plasma [47–50]. So, in this research, to find the physical insight into the synergistic enhancement in MF-CAP treatment, the purpose is to pay attention to the effect of the magnetic field on plasma characteristics, based on the changes of electron parameters in the presence of magnetic field including the various electron temperatures and the electron density.

The results of the electron density calculation show that a significant change ( $4.65 \times 10^{15} \, cm^{-3}$ to $4.095 \times 10^{16} \, cm^{-3}$ ) in the electron density is observed in the presence of a magnetic field, possibly created only by placing the magnet near the plasma. These values are in relatively good agreement with data reported in Refs [42–45]. It can be concluded that a significant part of the effects observed in previous works [6,18] may be attributed to the enhancement in the magnitude of electron density led to increased reactive species production.

Besides the particle densities, the importance of the plasma temperature among the plasma characteristics is widely expressed [46–48]. Determining the plasma temperature is crucial for characterizing the other plasma properties, such as the relative populations of energy levels and the particle velocity distribution [49]. It controls the effectiveness and the efficiency of plasma and is responsible for initiating various reactions in plasma systems [50]. Various electron energies correspond to the different active species' productivities. Hence, several reaction rates may be directly regulated by the electron temperature ($T_e$). Numerous reaction's rate coefficients depend on the energy and temperature of the electrons. Generally, the reactions not requiring the electron energies to "go" are unaffected by changes in the electron temperature. Nearly all reactions that required the electron energies to "go" tended to have their maximum values of the rate constants at the highest electron energies [51]. The density of negative ions in low-pressure hydrogen plasma is a function of electron temperature and neutral gas pressure. Actually, when the electron temperature drops, the population of negative ions decreases [52]. This effect is also shown for mixed plasma (Ar-methane) so that negative hydrogen ion production is controlled by electron temperature [53]. For these reasons, the important role of the electron temperature in cold plasma and its effect on various parameters, especially negative ions density, have been discussed in most papers on plasma applications [53–54] and many ways have been examined to control or increase the electron temperature [55].

According to figure 3, changes of 0.8 eV have been observed in the electron temperature in the presence of SMF. Therefore, it can be expressed that the observed effects in previous reports, for example, our previous results on the cell's viability and migration rate treated by the MF-CAP system, were caused by the increase of negative ions in the plasma and cell culture medium, due to increasing the electron temperature.

The gas temperature, a scale of the neutral particles' kinetic energy, is an effective component in plasma chemistry. So, its determination can be considered a goal in the diagnostics of most plasma applications. Inferring gas temperature from the rotational temperature of nitrogen molecules yielded a high rotational temperature. As we've treated cells without disturbing their health, this is a strong indication for the wrong assumption of the equality of $T_g$ and $T_{rot}$. Generally, the extreme regimes of the highly collisional high-pressure plasmas ensure rotational-translational relaxation, however, it has been discussed extensively that the equality of gas temperature to the rotational temperature is not universal. Accordingly, it demands to confirm the fulfillment of one of the following conditions, i.e., a thermalized rotational distribution or a fast rotational energy transfer for equilibrating the non-thermal rotational distributions [32]. The verification should be achieved beforehand to prevent a wrong estimation of the gas temperature from rotational temperature. Additionally, argon atoms might not be the sole influencer on excitation to the higher rotational levels, but the vibrational states of $N_2(C)$ impact on the rotational distribution [56].

A comparison of reported quantities in figure 6 a, b reveals the unchanged translational temperature under a magnetic field. It demonstrates that the kinetic energy of nitrogen molecules is independent of the magnetic field. Another specification that may provide information about the exchanges between particles is the vibrational temperature. Here, the translational-vibrational equilibrium manifests efficient energy transfer between vibrational and translation freedoms of the nitrogen molecules, regardless of the magnetic field. The vibrational translational (V-T) energy transfer has been studied in the subject of the plasma kinetics for modeling the relaxation effects in heating phenomena of molecular species. V-T energy transfer is usually slow, and the equilibrium is not expected as the rotational and translational energy conversion is more facile. A kinetic mechanism survey is needed to illustrate the excitation-relaxation competition in plasma [57–59].

The SMF has different effects on CAP, regarding its location in front of the plasma effluent. A magnetic field perpendicular to the discharge symmetry can cause more excitation and ionization due to the electron's curved movement in the presence of a magnetic field [60]. It may not mean there is higher electron temperature which is the presumed situation in the axial fields [61]. It can be expected that there is higher excitation temperature in the presence of a magnetic field which is confirmed by the temperature enhancement from 5950 K to 6500 K, as it is clearly shown in Figure 6.

6. **Conclusion**

The combined effect of cold atmospheric plasma and magnetic field has been investigated in the past to improve plasma performance; however, the main reason for this promotion was not apparent. In this research, the characteristics of RF plasma in the presence of a magnetic field are examined. The results showed that an external magnetic field perpendicular to the plasma plume increased the electron temperature and density. It should be mentioned that the enhancement of the plasma temperature is an important factor in many biological and medical applications. The simulation of the nitrogen optical spectrum by Specair software showed that the electronic excitational and rotational temperature increased, but the vibrational and translational temperature remained constant and equal under the magnetic field. However, results confirmed the thermalization criterion check requirement for estimating gas temperature from rotational distribution. Moreover, the result presented that the temperature and short reactive species (electron) incrimination are the main reason for plasma effect enhancement. Since this combination effect is most utilized in medical applications of plasma, temperature change should be considered to prevent thermal damage.

**Competing interests**
The authors declare no competing interests.

**Data availability statements**
The datasets generated or analyzed during the current study are available and presented just in this paper.